\documentclass[12pt,draftclsnofoot,onecolumn]{IEEEtran}
\IEEEoverridecommandlockouts

\usepackage{cite}
\usepackage{amsmath,amssymb,amsfonts}
\usepackage{algorithmic}
\usepackage{graphicx}
\usepackage{textcomp}
\usepackage{xcolor}
\usepackage{bm}
\usepackage{float}
\usepackage{siunitx}
\usepackage{url}    
\usepackage{standalone}            
\usepackage{color}
\usepackage{tikzscale}
\usepackage{pgfplots}
\usepackage{tikz}

\DeclareMathOperator{\Det}{Det}
\DeclareMathOperator{\diag}{diag}    
\DeclareMathOperator*{\argmax}{arg\,max}
\DeclareMathOperator*{\tr}{tr}
    
\usetikzlibrary{shapes.geometric}    
\usetikzlibrary{fit,shapes.geometric}
\usetikzlibrary{shapes,arrows,chains}
\pgfdeclarelayer{signal}
\pgfsetlayers{signal,main}
\tikzset{pics/.cd,
  SBS/.style={code={
      \begin{scope}[local bounding box=#1]
      \fill [pic actions/.try] (-1,0) -- (-1/2,3) -- (1/2, 3) -- (1,0) -- cycle;
      \fill [pic actions/.try] (-1/16,2) rectangle (1/16,4);
      \fill [pic actions/.try] (0,4) circle [radius=1/4];
      \foreach \i in {-1,1}
        \fill [shift=(90:4), xscale=\i]
          \foreach \r in {1,3/2,2}{
            (-45:\r) arc (-45:45:\r) -- (45:\r-1/10)
            arc(45:-45:\r-1/10) -- cycle
          };
       \end{scope}
  }},
  SU/.style={code={
      \begin{scope}[local bounding box=#1]
      \fill [pic actions/.try] (-5/2,3/2) -- (-5/2,3) -- (1/2, 3) -- (1/2,3/2) -- cycle;
      \fill [pic actions/.try] (-1/16,2) rectangle (1/16,4);
      \fill [pic actions/.try] (0,4) circle [radius=1/4];
      \fill [red,pic actions/.try] (-2,2) circle [radius=1/5];
      \fill [green,pic actions/.try] (-2,5/2) circle [radius=1/5];
      \fill [white,pic actions/.try] (-0.75,2.25) circle [radius=1/3];
      \foreach \i in {-1,1}
        \fill [shift=(90:4), xscale=\i]
          \foreach \r in {0.75,1}{
            (-45:\r) arc (-45:45:\r) -- (45:\r-1/10)
            arc(45:-45:\r-1/10) -- cycle
          };
       \end{scope}
  }},
  SIGNAL/.style={code={
    \begin{scope}[local bounding box=#1]
      \fill [pic actions/.try]
      (0,-3) -- (-1,1/2) -- (1/8,1/4) -- (0,3) -- (1,-1/2) -- (-1/8,-1/4)
      -- cycle;
    \end{scope}
  }}
}
\colorlet{sky blue}{blue!60!cyan!75!black}
\colorlet{dark blue}{blue!50!cyan}
\colorlet{chameleon}{olive!75!green}
\tikzset{signal/.style={draw=gray, line width=0.2em, dashed}}
    
\begin{document}

\title{Entropy-Based Sensing Schemes for Energy Efficiency in Massive MTC}
\author{Sergi~Liesegang,~\IEEEmembership{Student Member,~IEEE}, Antonio Pascual-Iserte,~\IEEEmembership{Senior Member,~IEEE}, and Olga Mu\~noz,~\IEEEmembership{Member,~IEEE}

\thanks{This work is supported by the project 5G\&B RUNNER-UPC (TEC2016-77148-C2-1-R (AEI/FEDER, UE)) and the research network RED2018-102668-T Red COMONSENS (funded by the Spanish Ministry of Science, Innovation and Universities), the fellowship FPI BES-2017-079994, and the grant 2017 SGR 578 (funded by the Catalan Government - AGAUR).}
\thanks{The authors are with the Department of Signal Theory and Communications,
Universitat Polit\`ecnica de Catalunya, Barcelona 08034, Spain (e-mail: sergi.liesegang@upc.edu, antonio.pascual@upc.edu, olga.munoz@upc.edu).

\copyright 2020 IEEE. Personal use of this material is permitted. Permission from IEEE must be obtained for all other uses, in any current or future media, including reprinting/republishing this material for advertising or promotional purposes, creating new collective works, for resale or redistribution to servers or lists, or reuse of any copyrighted component of this work in other works.

DOI: 10.1109/LWC.2020.2984223

}}

\markboth{Accepted Paper at IEEE Wireless Communications Letters (vol. 9, no. 8, Aug. 2020)}
{}

\maketitle

\begin{abstract}
Machine-type communications (MTC) are crucial in the evolution of mobile communication systems. Within this context, we  distinguish the so-called massive MTC (mMTC), where a large number of devices coexist in the same geographical area. In the case of sensors, a high correlation in the collected information is expected. In this letter, we evaluate the impact of correlation on the entropy of a set of quantized Gaussian sources. This model allows us to express the sensed data with the data correlation matrix. Given the nature of mMTC, these matrices may be well approximated as rank deficient. Accordingly, we exploit this singularity to design a technique for switching off several sensors that maximizes the entropy under power-related constraints. The discrete optimization problem is transformed into a convex formulation that can be solved numerically.
\end{abstract}


\section{Introduction}

Machine-type communications (MTC) have drawn a lot of attention during the past decade. They are key for the development of a plethora of applications in future mobile generations \cite{Sha15}. MTC systems involve transmissions between autonomous devices without human interaction. This facilitates the rapid deployment of large networks, which leads to an exponential growth in the number of connected terminals \cite{Daw17}. As a result, MTC are contributing to the advance of new fields of interest such as the Internet-of-Things \cite{Pal16}.

In this letter, a scenario with a large number of sensors is analyzed. These systems are commonly denoted as \textit{massive} MTC (mMTC) \cite{Boc16} and, given the high spatial density of these devices, the sensed data can be highly correlated \cite{Sha13}. Thereby, a natural question that arises is how to manipulate this information to reduce the overall payload. In fact, this data redundancy entails the waste of a significant amount of power. Hence, working with only the relevant information would help increase the energy efficiency and extend the battery lifetime of the sensors, which are critical issues in MTC deployments.

To tackle this problem, we formulate strategies to decide that some devices will not sense while still ensuring that a significant  amount of the information is collected \cite{Row07,Jos09,Che15}. The procedure is based on the entropy of quantized Gaussian sources, which allows us to represent the collected information through the (frequently ill-conditioned) data correlation matrix \cite{Sca05,Dab10}. Accordingly, since this matrix captures the correlation between the sensors data, we can decide which sensors will be active and which will remain silent, thus reducing the energy consumption. In particular, we will select as active devices those that maximize the information collected (i.e., entropy) under power-related constraints and silence the rest. Since devices transmit only after sensing, our approach can further reduce the power consumption. To evaluate the perfor-mance of these schemes, we study the loss in the data entropy after the device selection. To the best of our knowledge, no studies have been reported in this direction so far.

Given the discrete nature of the optimization problem (i.e., sensor selection), convex relaxation and approximations of the entropy are used to obtain sub-optimal solutions. We show the new formulated problem to be convex and, thus, numerically solvable with standard optimization methods. The performance of the proposed strategy is evaluated in an illustrative practical scenario in which the elements of the correlation matrix are generated through a simple spatial model based on the distance between sensors. Under certain conditions, this matrix can be shown to be well approximated by a singular matrix.

This letter is structured as follows. In Section II, the system model, the data entropy and the correlation are described. In Section III, the sensing scheme is formulated and the proposed solution is presented. Section IV is devoted to the numerical simulations and conclusions are shown in Section V. 
 
\section{System Model}

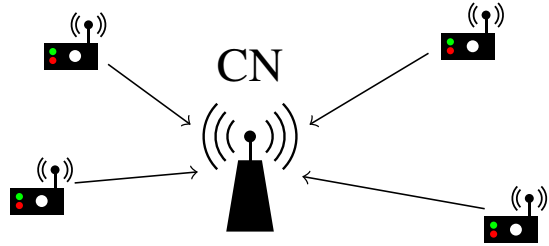
\begin{figure}[t]
\centerline{\begin{tikzpicture}[x=1em,y=1em]

\begin{scope}[local bounding box=b1]
\pic [x=1em*4/8, y=1em*4/8,line width=0.05em,fill=black] at (0,-2.5) {SBS=p1};
\pic [x=1em*3/8, y=1em*3/8,line width=0.05em, fill=black]  at (13.2259,4.8087)  {SU=q2};
\pic [x=1em*3/8, y=1em*3/8,line width=0.05em, fill=black]  at (15.6493,-5.4736)  {SU=q3};
\pic [x=1em*3/8, y=1em*3/8,line width=0.05em, fill=black]  at (-9.0517,4.2616)  {SU=q7};
\pic [x=1em*3/8, y=1em*3/8,line width=0.05em, fill=black]  at (-10.9251,-3.8705)  {SU=q6};

\node [] at (0,2.25) {$\textrm{CN}$};
\draw [<-] (1.25,1) -- node [pos = 0.9,below=0.1em]{} (3.75,2.5);
\draw [<-] (p1) -- node [pos = 0.6,below=0.001em]{} (q3);
\draw [<-] (p1) -- node [pos = 0.7,below=0.1em]{} (q6);
\draw [<-] (-1.25,1) -- node [pos = 1,below=0.1em]{} (q7);
\end{scope}

\end{tikzpicture}}
\caption{System setup for $M = 4$.}
\label{fig:1}
\end{figure}

Throughout this letter, we consider a setup with $M$ sensors deployed around one data collector node (CN), as depicted in Fig.~\ref{fig:1}. In this scenario, we focus on how much information the sensed data contains, and leave the communication aspects for future studies. Then, it is desirable to establish a metric that quantifies this information. It is natural to use the multivariate entropy as a proper indicator, since it determines the minimum number of bits needed to represent a discrete source \cite{Cov06}. 

To describe the sensed data, we define a random vector $\bm{X} = [X_1,\ldots,X_M] \in \mathbb{R}^M$ with distribution $f_{\bm{X}}(\bm{x})$, where each element represents the information collected by each sensor. Since this information is continuous, an infinite number of bits are needed to represent $\bm{X}$. Thus, it is necessary to quantize the source to work with finite precision. We denote the quantized version as $\bm{X}^{\Delta}$ with associated mass probability distribution $p_{\bm{X}^{\Delta}}(\bm{x}^{\Delta})$. The multivariate entropy (in bits) is
\begin{equation}
H(\bm{X}^{\Delta}) = - \sum_{\bm{x}^{\Delta}} p_{\bm{X}^{\Delta}}(\bm{x}^{\Delta}) \log_2 p_{\bm{X}^{\Delta}}(\bm{x}^{\Delta}),
\label{eq:1}
\end{equation}
and, as shown in the Appendix, it can be lower bounded by
\begin{equation}
\widetilde{H}(\bm{X}^{\Delta}) \triangleq h(\bm{X}) - M \log_2 \Delta \leq H(\bm{X}^{\Delta}),
\label{eq:2}
\end{equation}
where we have considered a uniform quantization step $\Delta$. The term $h(\bm{X})$ is the differential entropy of the continuous source:
\begin{equation}
h(\bm{X}) = - \int_{\bm{x}} f_{\bm{X}}(\bm{x}) \log_2 f_{\bm{X}}(\bm{x}) d \bm{x}.
\label{eq:3}
\end{equation}

Note that the previous lower bound $\widetilde{H}(\bm{X}^{\Delta})$ becomes tight for $\Delta \to 0$ \cite{Cov06}, as shown in Section~\ref{sec:5}. This is why it will be used as an approximation for the actual entropy $H(\bm{X}^{\Delta})$.

Considering the Gaussian distribution with zero mean and (full-rank) correlation matrix $\bm{C} \in \mathbb{R}^{M \times M}$, i.e., $\bm{X} \sim \mathcal{N} (\mathbf{0},\bm{C})$, we have \cite{Cov06}:
\begin{equation}
h(\bm{X}) = \frac{1}{2} \left(M \log_2 2 \pi e + \log_2 \det (\bm{C})\right).
\label{eq:4}
\end{equation}

Note that the Gaussian distribution is taken as reference as it requires the largest number of bits for accurate representation once quantized (hence, it can be understood as a challenging case) \cite{Cov06}. Furthermore, the Gaussian assumption is typical in many sensing scenarios \cite{Sca05,Dab10}. Also, it allows us to measure the amount of collected information as a function of matrix $\bm{C}$, which models the correlation between sensors data.

The elements of $\bm{C}$ can be constructed as follows \cite{Dab10}:
\begin{equation}
[\bm{C}]_{i,j} = \sigma_i \sigma_j K(d_{i,j}), \quad 1 \leq i \leq M, \quad 1 \leq j \leq M,
\label{eq:5}
\end{equation}
where $\sigma_i$ is the standard deviation of the information signal collected by sensor $i$. The term $K(d_{i,j})$ is the correlation factor between the data of sensors $i,j$ with $\vert K(d_{i,j}) \vert \leq 1$, and that usually depends on the distance between them $d_{i,j}$. Different models can be adopted for the function $K(d_{i,j})$ \cite{Ber01}. Just as an illustrative example, a well-known and accepted model for sensor networks is a quadratic-distance exponential \cite{Vur04}:
\begin{equation}
K(d_{i,j}) = \textrm{exp}(- (d_{i,j}/\theta)^2),
\label{eq:6}
\end{equation}
where $\theta > 0$ controls the range of correlation. The extreme values in \eqref{eq:6} are $0$ and $1$ for $d_{i,j} = \infty$ and $d_{i,j} = 0$, respectively. Hence, as devices are located at different positions, the spatial dependency of the correlation is incorporated in the model.

Note that the strategy developed in this letter is based on a generic matrix $\bm{C}$ and is also valid if other underlying models for matrix $\bm{C}$ are adopted (see examples \cite{Sha13,Sca05,Dab10,Vur04}).

Under some assumptions, the correlation matrix $\bm{C}$ can be well approximated by a singular matrix for $M \to \infty$ \cite{Sca05}. Intuitively, this can be understood by the fact that its elements depend only on the distance and, given the high spatial density, their values can be very similar. From now on, and to ease notation, $\bm{C}$ will represent the singular (rank deficient) matrix approximating the original correlation matrix. The differential entropy of the resulting degenerate Gaussian distribution is \cite{Sca05}
\begin{equation}
h(\bm{X}) = \frac{1}{2} \left(r({\bm{C}}) \log_2 2 \pi e + \log_2 \Det (\bm{C})\right),
\label{eq:7}
\end{equation}
where $r({\bm{C}})$ and $\Det (\bm{C})$ are the rank and pseudo-determinant of $\bm{C}$, respectively \cite{Kni14}. The latter is computed as the product of the non-zero eigenvalues of $\bm{C}$. Finally, the entropy of the quantized vector $\bm{X}^{\Delta}$ yields \cite{Sca05}
\begin{equation}
H(\bm{X}^{\Delta}) \geq h(\bm{X}) - r(\bm{C}) \log_2 \Delta = \widetilde{H}(\bm{X}^{\Delta}).
\label{eq:8}
\end{equation}

\section{Entropy Based Sensor Selection}

\subsection{Problem Formulation}

The purpose of this letter is to present a strategy that minimizes the loss of information when silencing some sensors. For this task, we take advantage of the high correlation in the sensed data, which yields an ill-conditioned correlation matrix $\bm{C}$, and select as active sensors those with the highest impact on $\bm{C}$. Note that as matrix $\bm{C}$ is singular, the number of sensors to keep active does not increase substantially with $M$ \cite{Sca05}.

We are interested in designing a device selection matrix $\bm{B} = \diag(\bm{b}) = \diag([b_1, \ldots, b_M])$, with binary elements $b_i \in \{0,1\}$, that reduces power consumption while preserving the sensed information to be transmitted. It represents an on/off activity strategy applied to the sensors. The reason behind this hard allocation scheme is the simplicity of sensors, which may be unable to handle various sensing states (they can be either active or asleep, but not in intermediate phases). 

Let us denote $\bm{X}_p^{\Delta}$ as the quantized version of the selected data $\bm{X}_p = \bm{B} \bm{X}$. Then, the goal is to find a selection matrix $\bm{B}$ that maximizes $H(\bm{X}_p^{\Delta})$ under a norm constraint. To that end, we consider the following optimization problem:
\begin{equation}
\bm{b}^{\star} = \underset{\bm{b} \in \{0,1\}^M}{\argmax} \, H(\bm{X}_p^{\Delta}) \quad \textrm{s.t.} \quad \| f(\bm{b}) \| \leq \delta,
\label{eq:9}
\end{equation}
where $\| f(\bm{b}) \|$ represents a power-related constraint. The constant $\delta > 0$  determines the bound applied to the norm. 

Furthermore, since the correlation matrix of $\bm{X}_p^{\Delta}$, i.e., $\bm{BCB}$, may also be rank deficient, the corresponding approximate entropy of this new quantity is
\begin{equation}
\widetilde{H}(\bm{X}_p^{\Delta}) \triangleq h(\bm{X}_p) - r(\bm{BC}\bm{B}) \log_2 \Delta \leq H(\bm{X}_p^{\Delta}),
\label{eq:10}
\end{equation}
with differential entropy
\begin{equation}
h(\bm{X}_p) = \frac{1}{2} \left(r(\bm{BC}\bm{B}) \log_2 2 \pi e + \log_2 \Det (\bm{BC}\bm{B})\right).
\label{eq:11}
\end{equation}

Given that the lower bound $\widetilde{H}(\bm{X}_p^{\Delta})$ becomes tight for small $\Delta$, we can use it as the objective function in \eqref{eq:9}. The exact error of this approximation will be illustrated in Section~\ref{sec:5}. When maximizing the lower bound, the problem in \eqref{eq:9} becomes
\begin{equation}
\bm{b}^{\star} = \underset{\bm{b} \in \{0,1\}^M}{\argmax} \, \widetilde{H}(\bm{X}_p^{\Delta}) \quad \textrm{s.t.} \quad \| f(\bm{b}) \| \leq \delta.
\label{eq:12}
\end{equation}

Regarding the constraint function $\| f(\bm{b}) \|$, we can use different approaches. For instance, some reasonable options are:
\begin{enumerate}
	\item[(i)] Number of active sensors: $\| \bm{b} \|_0 $,
	\item[(ii)] Sum of powers: $\| \bm{b}^2 \circ \bm{\gamma}  \|_1 $,
	\item[(iii)] Maximum power: $\| \bm{b}^2 \circ \bm{\gamma}  \|_{\infty} $,
\end{enumerate}
where $\circ$ is the element-wise or Hadamard product and $\bm{\gamma}  = [\gamma_1, \ldots, \gamma_M]$ is the set of individual sensor power consumptions. When applied to \eqref{eq:12}, (i) determines the sparsity of $\bm{b}$ and, although it does not consider the power $\bm{\gamma}$, it can reduce the consumption since it limits the number of active sensing devices. On the other hand, since (ii) and (iii) consider power consumption, both can improve the system energy efficiency better than (i). However, since $\bm{\gamma}$ is usually fixed, (iii) has no special interest as the solution would be a trivial thresholding irrespective of $\widetilde{H}(\bm{X}_p^{\Delta})$, i.e., $b_i = 1$ for all sensors with power $\gamma_i$ below $\delta$. This is why we will focus on (i) and (ii).

Overall, given the binary nature of $\bm{b}$, we formulate a lossy information approach that prioritizes the relevant components of the sensed data (i.e., those with more information). This way, we end up with a system that delivers the most relevant information with the benefit of a reduced  consumption, i.e., the battery lifetime can be extended with similar performance.

Unfortunately, the solution to problem \eqref{eq:12} is combinatorial and might be unfeasible to solve for large $M$. This is the reason why, in the upcoming section we propose some simplifications to reduce computational cost.

\subsection{Proposed Solution} \label{sec:4}

To reduce the problem to a feasible complexity, in this letter we take the following steps: relaxation of the constraint set, approximation of the objective function, and rounding of the obtained (sub-optimal) solution. This way, we end up with a convex optimization problem, easy to solve with standard numerical methods, while still providing good performance.

We start by substituting the discrete constraint on $\bm{b}$, i.e., $\bm{b} \in \{0,1\}^M$, with a convex relaxation. In particular, now $\bm{b}$ can take continuous values between $0$ and $1$, i.e., $\bm{b} \in [0,1]^M$. Thereby, we can express the new continuous problem as
\begin{equation}
\bm{b}^{\star} = \underset{\bm{b} \in [0,1]^M}{\argmax} \, \widetilde{H}(\bm{X}_p^{\Delta}) \quad \textrm{s.t.} \quad \| f(\bm{b}) \| \leq \delta.
\label{eq:13}
\end{equation}

Note that the $\log_2 \Det$ function in $\widetilde{H}(\bm{X}_p^{\Delta})$ is not concave in $\bm{B}$ as the $\Det$ function is discontinuous \cite{Kni14}. This is why, in order to deal with proper determinants, we make use of the Cholesky decomposition of the correlation matrix \cite{Pet12}. Since $\bm{C}$ is symmetric and positive semi-definite ($\bm{C} \succeq \mathbf{0}$), there exists a lower triangular matrix $\bm{L} \in \mathbb{R}^{M \times M}$ such that $\bm{C} = \bm{L}\bm{L}^{\textrm{T}}$. Note that for rank-deficient $\bm{C}$, the matrix $\bm{L}$ will have $M - r(\bm{C})$ columns equal to zero, i.e.
\begin{equation}
\bm{L} = \begin{bmatrix}
    l_{11}       & 0      & 0 	   & \ldots 		& 0	        & \ldots & 0      \\
    l_{21}       & l_{22} & 0      & \ldots			& 0  	    & \ldots & 0  \\
    \vdots       & \vdots & \ddots & \ddots         & \vdots  	&        & \vdots \\
    l_{M1}       & l_{M2} & \ldots & l_{Mr(\bm{C})} & 0         & \ldots & 0
\end{bmatrix} = [ \bm{L}_r \, \vert \,  \mathbf{0} ],
\label{eq:14}
\end{equation}
where $\bm{L}_r \in \mathbb{R}^{M \times r(\bm{C})}$ is the reduced version of $\bm{L}$. With that, the correlation matrix can be expressed as $\bm{C} = \bm{L}_r\bm{L}_r^{\textrm{T}}$. In addition, given that the pseudo-determinant is also translation invariant \cite{Kni14}, we have that
\begin{equation}
\Det \bm{C} = \Det (\bm{L}_r\bm{L}_r^{\textrm{T}}) = \Det (\bm{L}_r^{\textrm{T}}\bm{L}_r) = \det (\bm{L}_r^{\textrm{T}}\bm{L}_r).
\label{eq:15}
\end{equation}

Similarly, we can write $\Det (\bm{BC}\bm{B}) = \Det (\bm{L}_r^{\textrm{T}} \bm{B} \bm{B} \bm{L}_r)$, which becomes a proper determinant if the selection matrix is positive definite, i.e., $\bm{B} \succ \mathbf{0}$. This way, the $\log_2 \Det$ in $\widetilde{H}(\bm{X}_p^{\Delta})$ yields the (continuous) $\log_2 \det$. However, the $\log_2 \det $ function is still not concave in $\bm{B}$. In what follows, we use instead matrix $\bm{P}$, defined as $\bm{P} = \diag(\bm{p}) = \bm{B} \bm{B} = \bm{B}^2$, and reformulate the problem in \eqref{eq:13} as follows:
\begin{equation}
\bm{p}^{\star} = \underset{\bm{p} \in [0,1]^M}{\argmax} \, \widetilde{H}(\bm{X}_p^{\Delta}) \quad \textrm{s.t.} \quad \| f(\bm{p}) \| \leq \delta,
\label{eq:16}
\end{equation}
where now, defining $\bm{\Gamma} = \diag(\bm{\gamma})$ and  neglecting (iii), the constraint functions in $\bm{p}$ are given by:
\begin{itemize}
	\item[(i)] Number of active sensors: $\| \bm{p} \|_0 $,
	\item[(ii)] Sum of powers: $\| \bm{p} \circ \bm{\gamma}  \|_1 = \tr(\bm{P\Gamma})$.
\end{itemize}

As a second step, we substitute the rank in the objective function $\widetilde{H}(\bm{X}_p^{\Delta})$ in \eqref{eq:16} by its nuclear norm \cite{Igl18}, producing a lower bound of the objective function that is to be maximized and that can be shown to be concave in $\bm{P}$ for $\bm{P} \succ \bm{0}$. Refer to the Appendix for more details on the derivation of this lower bound, denoted by $\widetilde{H}_{lb}(\bm{X}_p^{\Delta})$. In addition, we also substitute the $l_0$ quasi-norm in (i) by its closest convex approximation, namely the $l_1$ norm $\| \bm{p} \|_1 = \tr(\bm{P})$ \cite{Tro06}.

Note that the condition $\bm{P} \succ \mathbf{0}$ can be ensured by imposing $\bm{P}- \nu \mathbf{I} \succeq \mathbf{0}$ for a sufficiently small $\nu > 0$, which is equivalent to $p_i \geq \nu \ \forall i$. In fact, since we eventually round the solution, $p_i = \nu $ or $p_i = 0$ would lead to the same binary value (i.e., the same final solution). This practice is widely used in the area of sparse sensing (see \cite{Che16} and references therein).

As shown in the Appendix, thanks to the previous steps, we end up with the following convex problem (maximization of a concave function subject to convex constraints), where the optimization variable is the diagonal matrix $\bm{P}$:
\begin{equation}
\bm{P}^{\star} = \underset{\bm{P} = \diag(\bm{p}), \ \bm{p} \in [\nu,1]^M }{\argmax} \, \widetilde{H}_{lb}(\bm{X}_p^{\Delta}) \quad \textrm{s.t.} \quad \| f(\bm{P}) \| \leq \delta.
\label{eq:17}
\end{equation}

Note that, due to the determinant, an analytic closed-form solution is hard to find, even for small $M$. This is the reason why, given the convexity of the problem, standard optimization methods are used to find the global solution numerically \cite{Boy09}. More specifically, in order to solve \eqref{eq:17}, we use the successive approximation method with the SDPT3 solver of the CVX software package, and we set the precision to `high' \cite{CVX14}. 

Finally, once the solution is found in the continuous domain, we need to return to the discrete regime. To do so, we threshold (round) the solution such that the constraints are satisfied, i.e., activate sensors with largest $p_i^{\star}$ until (i) or (ii) are fulfilled.

\section{Numerical Results} \label{sec:5}

In this section, several simulations are presented to illustrate the performance of the sensing scheme. More specifically, the relative loss $\varepsilon$ in information after the device selection is evaluated for different sweeps where the parameters $\delta$ and $M$ are changed. This metric corresponds to the difference between the entropies of the quantized data defined in \eqref{eq:8} and \eqref{eq:10}, i.e., 
\begin{equation}
\varepsilon = \vert \widetilde{H}(\bm{X}^{\Delta}) - \widetilde{H}(\bm{X}_p^{\Delta}) \vert / \vert \widetilde{H}(\bm{X}^{\Delta}) \vert.
\end{equation}

Regarding the scenario, we consider sensors to be normally distributed in a small region around the CN. This small area reduces the computational complexity of the numerical methods, as we can use smaller values of $M$ while preserving the high spatial density of sensors.  As an example of application, we consider temperature as the phenomenon to be measured, and we make use of the quadratic-distance exponential model in \eqref{eq:6} with $\sigma_i = \sigma = \ang{50}$ and $\theta = 3.08$ for the correlation matrix $\bm{C}$. In addition, we also consider sensors to have different levels of power consumption, i.e., $\gamma_i \in [0.01, 0.2]$ W \cite{Nok16}. 

\begin{figure}[t]
\centerline{\includegraphics[scale = 0.75]{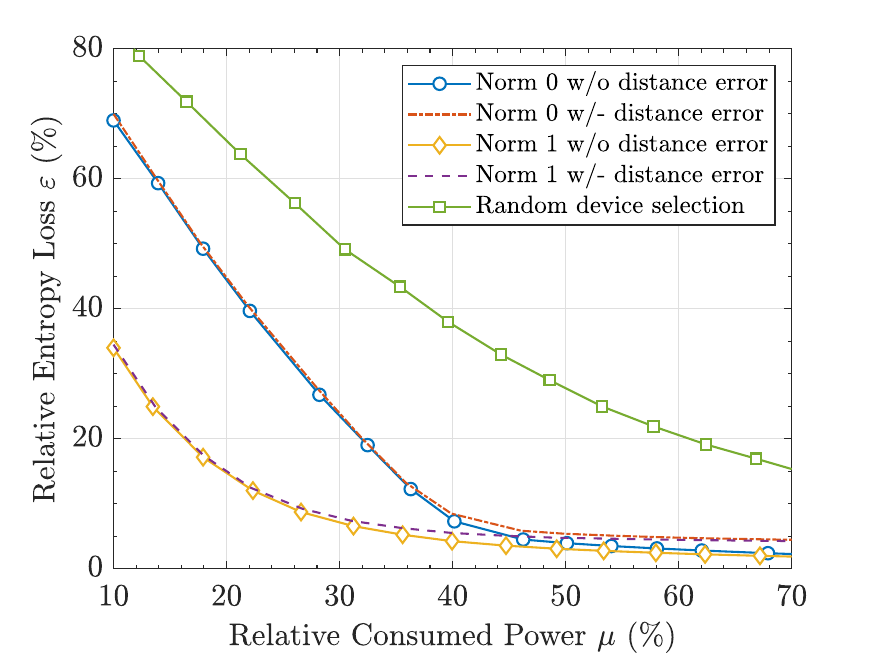}}
\caption{Relative entropy loss $\varepsilon$ versus consumed power $\mu$ with $M = 50$.}
\label{fig:2}
\end{figure}

\begin{figure}[t]
\centerline{\includegraphics[scale = 0.75]{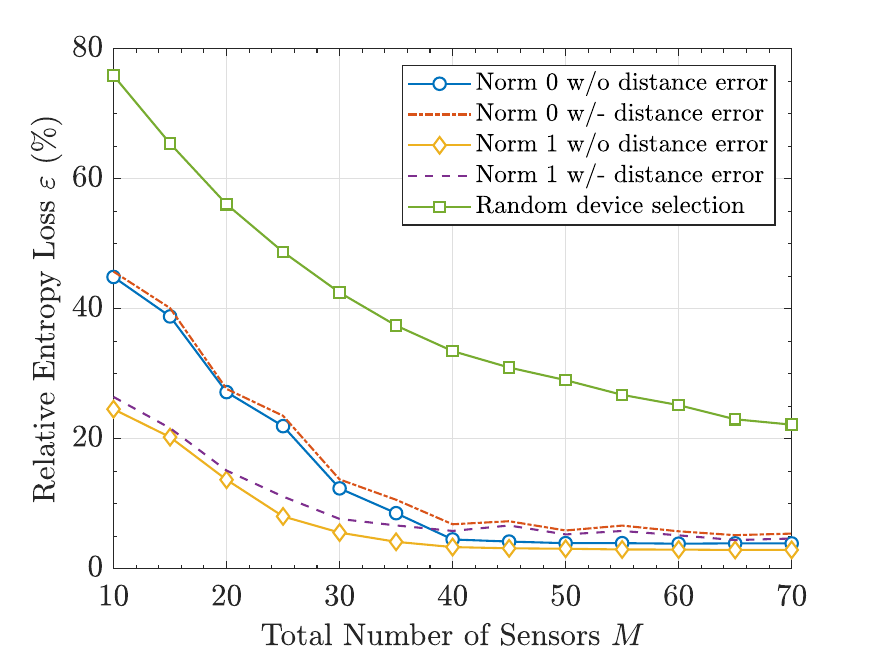}}
\caption{Relative entropy loss $\varepsilon$ versus $M$ with $\alpha = 0.5$.}
\label{fig:3}
\end{figure}

The sweeps over $\delta$ and $M$ are depicted in Fig.~\ref{fig:2} and Fig.~\ref{fig:3}, respectively. Note that the values of $\delta$ represent a percentage of the different constraint functions applied directly to $\bm{\Gamma}$, i.e., (i) $\delta = \alpha \tr(\mathbf{I}) = \alpha M$ and (ii) $\delta = \alpha \tr(\bm{\Gamma})$, with $\alpha \in [0,1]$. Then, $\alpha = 0.5$ is used in Fig.~\ref{fig:3} and $M = 50$ is set in Fig.~\ref{fig:2}. In addition, we set $\Delta$ according to the Appendix with $\nu = 10^{-3}$.

In order to actually represent the consumed power after the selection, in Fig. \ref{fig:2} we plot the relative entropy loss $\varepsilon$ with respect to (w.r.t.) the relative consumed power $\mu = \tr(\bm{P\Gamma})/\tr(\bm{\Gamma})$ instead of $\alpha$. In fact, note that $\alpha$ is equal to $\mu$ in constraint (ii) while it is indirectly related to this ratio in the case of (i).

Also, since in real scenarios the exact distances might not be known, but estimated, in Fig. \ref{fig:2} and Fig. \ref{fig:3} we show the entropy loss when the distances are perfectly known and also for an additive Gaussian estimation error with zero-mean and standard deviation of $10\%$ w.r.t. to the actual distances.

As observed in Fig.~\ref{fig:2}, higher values of $\mu$ ($\alpha$) yield smaller entropy losses. This is not surprising as we allow more sensors to be active. For instance, for $\mu = 40\%$ we have a relative loss of $7.30\%$ and $4.25\%$ for constraints (i) and (ii). This shows that it is possible to reduce significantly the consumed power while preserving most of the relevant information.

In Fig. \ref{fig:3} we can see that the entropy loss diminishes with the total number of sensors $M$ because the dimension of the null-space of $\bm{C}$ increases with $M$. For an increasing $M$ and a constant $\alpha$, a significant number of sensors can be silent while reducing the entropy loss. A large error $\varepsilon$ is then obtained for small $M$, when the correlation matrix is full-rank (all sensors provide relevant information). For instance, for $M=10$ the losses are around $45\%$ and $25\%$ for constraints (i) and (ii), whereas for $M=40$, the entropy loss is only $4.5\%$ and $3.3\%$.

Note that a better performance is attained for constraint (ii) because the optimal sensors (i.e., those with the highest impact on $\bm{C}$ and smaller consumption) can also be distinguished in terms of consumed energy. Hence, the difference in power consumption can play an important role in the optimization as more relevant information is taken into account in the problem. 

To compare the performance of our system, the relative loss of a random selection is also illustrated in Fig.~\ref{fig:2} and Fig.~\ref{fig:3}. It represents the case where sensors are randomly chosen until constraint (ii) is fulfilled. As expected, a higher loss is obtained with this naive approach, specially for small $\alpha$ and $M$, where the optimal solution (i.e., device selection) is less flexible. 

In contrast to the approaches described in \cite{Jos09,Che15}, where the sensor selection is based on the estimation of the  parameter to be measured, we rely on the correlation between measurements (represented with the matrix $\bm{C}$). Thus, the selection is based on this statistical information. In addition, since the optimization procedure is done at the CN and requires low signalling (i.e., binary messages to report the on/off decisions), there is no significant additional power consumption of the sensors.

\begin{figure}[t]
\centerline{\includegraphics[scale = 0.75]{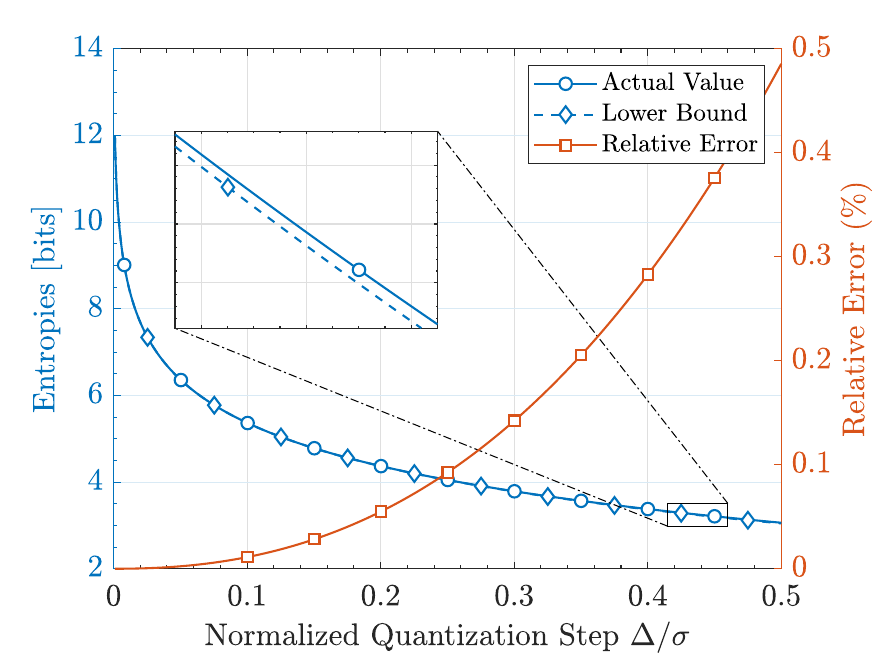}}
\caption{Actual entropy and lower bound (left) and relative error (right) versus normalized quantization step $\Delta/\sigma$ with $M = 1$.}
\label{fig:4}
\end{figure}

To justify the approximation of the original entropy $H(\bm{X}^\Delta)$ by its lower bound $\widetilde{H}(\bm{X}^\Delta)$, both magnitudes, jointly with the relative error, are illustrated w.r.t. $\Delta/\sigma$ in Fig.~\ref{fig:4} with $M = 1$ ($\bm{X}^{\Delta} = X^{\Delta}$). The values of $H(X^{\Delta})$ are obtained numerically through simulations. As shown in the figure, the approximation is very tight, specially for low values of $\Delta/\sigma$. 

Finally, some practical issues should be pointed out. First, as we observe in Fig. \ref{fig:2} and Fig. \ref{fig:3}, our method is robust to the uncertainty in the distance estimation since the resulting entropy loss does not increase substantially. Second, as reducing the number of active devices reduces the number of transmissions (sensors will transmit if they collect information), our strategy can reduce the number of collisions as well.

\section{Conclusions}

In this letter, we have designed a selection strategy in dense MTC networks that minimizes the loss of information to be transmitted (i.e., sensed data entropy) subject to power-related constraints, i.e., minimize the loss in information when some of the sensors do not collect or report the sensed information. A degenerate Gaussian distribution is used as a model to measure the correlated sensed information. Given the discrete nature of the resulting optimization problem, some steps are taken to find a sub-optimal solution. Simulation results prove that most of the information (i.e., entropy) can be retrieved with a smaller number of active devices. This effect is more pronounced in mMTC, where a higher correlation is experienced. 

\section*{Appendix}

\subsection*{Entropy of Quantized Source}

According to \cite{Gal08}, the entropy of a quantized source $\bm{X}^{\Delta}$ already defined in \eqref{eq:1} can also be written in the following way:

\begin{equation}
H(\bm{X}^{\Delta}) = - \int_{\bm{x}} f_{\bm{X}}(\bm{x}) \log_2 \bar{f}(\bm{x}) d \bm{x} - M\log_2 \Delta,
\label{eq:19}
\end{equation}
where $\bar{f}(\bm{x})$ is the average value of $f_{\bm{X}}(\bm{x})$ over each quantization interval. Therefore, it can be seen that \eqref{eq:19} can be lower bounded by \eqref{eq:2}. In fact, this bound is tight for $\Delta \to 0$, i.e., when $f_{\bm{X}}(\bm{x}) \approx \bar{f}(\bm{x})$ is valid (high rate assumption).

For simplicity, we continue with the one-dimensional case, although the same discussion holds for higher dimensions. In particular, for $M = 1$ the difference between both entropies, now $H(X^{\Delta})$ and $\widetilde{H}(X^{\Delta})$, is always non-positive. The proof follows from Gibbs' inequality $\ln x \leq x - 1$:
\begin{align}
\widetilde{H}(X^{\Delta}) - H(X^{\Delta}) &= \int_x f_{X}(x) \log_2 \bar{f}(x)/f_{X}(x) d x \nonumber \\
&\leq \left( \int_x f_{X}(x) (\bar{f}(x)/f_{X}(x) - 1) d x \right) \log_2 e \nonumber \\
&= \left( \int_x \bar{f}(x)  d x - \int_x f_{X}(x) d x \right) \log_2 e \nonumber \\
&= (1 - 1) \log_2 e = 0.
\label{eq:20}
\end{align}

\subsection*{Convexity of Optimization Problem}

In order to derive the convex problem in $\bm{P}$, i.e., \eqref{eq:17}, starting from \eqref{eq:16}, we need to find a new concave objective function and all constraints must be convex. Note that as we are dealing with norm-constraints, they are convex. For the $l_0$ quasi-norm, convexity is preserved through the $l_1$ norm relaxation. In addition, the constraint $\bm{p} \in [0,1]^M$ defines a convex set. The same holds for $\bm{p} \in [\nu,1]^M$, which is forced with $\bm{P} - \nu \bm{I} \succeq \mathbf{0}$. Recall that the latter condition implies $\bm{P} \succ \mathbf{0}$, which does not define a convex cone as the origin is not included \cite{Boy09}.

The objective function $\widetilde{H}(\bm{X}_p^\Delta)$ in \eqref{eq:16} can be approximated by a lower bound. To that end, we bound the rank function $r(\bm{X})$ of a symmetric positive semi-definite matrix $\bm{X}$ by its convex envelope, i.e., the nuclear norm $\| \bm{X} \|_* = \tr(\sqrt{\bm{X}^{\textrm{T}}\bm{X}}) \footnotemark[1] = \tr(\bm{X})$, as discussed in \cite{Igl18}. This lower bound is valid if the spectral norm (maximum eigenvalue) of $\bm{X}$ is smaller than $1$, i.e., $\|\bm{X}\|_2 =\lambda_{\textrm{max}}(\bm{X}) \leq 1$. Note that $\|\bm{X}\|_*$ and $\|\bm{X}\|_2$ are two different norms and can take different values.

\footnotetext[1]{The matrix square root operation $\sqrt{\bm{Y}}=\bm{X}$ implies that $\bm{Y}=\bm{X}^2$. Note that although this operation is different from the element-wise square root operation, both operations coincide when applied to diagonal matrices.}

By defining the normalized matrix $ \overline{\bm{C}}$ as $\overline{\bm{C}}= \bm{C} / \| \bm{C} \|_2$, we ensure that $\|\sqrt{\bm{P}}\, \overline{\bm{C}}·\sqrt{\bm{P}} \|_2 \leq 1 $. This comes from the fact that $\|\bm{P}\|_2 \leq 1$ and that for two symmetric positive semi-definite matrices $\bm{A}$ and $\bm{B}$, it can be proved that $\| \bm{AB} \|_2 \leq \| \bm{A} \|_2 \| \bm{B} \|_2$ \cite{Wan93}. Therefore, given that $r (\sqrt{\bm{P}}·\bm{C}·\sqrt{\bm{P}}) = r(\sqrt{\bm{P}} \,·\overline{\bm{C}}·\sqrt{\bm{P}})$, the objective function $\widetilde{H}(\bm{X}_p^\Delta)$ in \eqref{eq:16} can be lower bounded as:
\begin{align}
\widetilde{H}(\bm{X}_p^{\Delta}) &= \frac{1}{2} \left(r\left(\sqrt{\bm{P}}\, \bm{C}\sqrt{\bm{P}}\right) \log_2 \left(\frac{2\pi e}{\Delta^2}\right) + \log_2 \det \left(\bm{L}_r^{\textrm{T}} \bm{P} \bm{L}_r \right) \right) \nonumber \\
&= \frac{1}{2} \left(r\left(\sqrt{\bm{P}}\, \overline{\bm{C}}\sqrt{\bm{P}}\right) \log_2 \left(\frac{2\pi e}{\Delta^2}\right) + \log_2 \det \left(\bm{L}_r^{\textrm{T}} \bm{P} \bm{L}_r \right) \right) \nonumber \\
&\geq \frac{1}{2} \left( \|\sqrt{\bm{P}}\,\overline{\bm{C}}\sqrt{\bm{P}}\|_* \log_2 \left(\frac{2\pi e}{\Delta^2}\right) + \log_2 \det \left(\bm{L}_r^{\textrm{T}} \bm{P} \bm{L}_r \right) \right) \nonumber \\
&\stackrel{\textrm{(a)}}{=} \frac{1}{2} \left(\tr(\bm{P} \overline{\bm{C}}) \log_2 \left(\frac{2\pi e}{\Delta^2}\right) + \log_2 \det (\bm{L}_r^{\textrm{T}} \bm{P} \bm{L}_r) \right) \nonumber \\
&\triangleq \widetilde{H}_{lb}(\bm{X}_p^{\Delta}),
\label{eq:21}
\end{align}
where (a) follows from the definition of nuclear norm and the fact that $\tr(\sqrt{\bm{P}}\,\overline{\bm{C}}\sqrt{\bm{P}}) = \tr(\bm{P} \overline{\bm{C}})$, which results from the trace property: $\tr(\bm{A}\bm{B}) = \tr(\bm{B}\bm{A})$ for any matrices $\bm{A}$ and $\bm{B}$. 

Note that the lower bound in \eqref{eq:21} is concave in $\bm{P}$ because $\tr(\bm{P}\overline{\bm{C}})$ is linear in $\bm{P}$ and $\log_2 \det (\bm{L}_r^{\textrm{T}} \bm{P} \bm{L}_r)$ is concave in $\bm{P}$ for $\bm{P} \succ \bm{0}$ \cite{Boy09}. Thus, replacing the objective function in \eqref{eq:16} by its lower bound, we obtain the convex problem \eqref{eq:17}: maxi-mization of a concave function subject to convex constraints.

Besides, for our analysis to be consistent, the approximation in \eqref{eq:8} and \eqref{eq:10} must be accurate. To that end, the quantization interval $\Delta$ must be close to the minimum eigenvalues $\lambda_{\textrm{min}}(\bm{C})$ and $\lambda_{\textrm{min}}(\bm{PC})$. For symmetric positive semi-definite matrices, we have that $\lambda_{\textrm{min}}(\bm{PC}) \geq \lambda_{\textrm{min}}(\bm{P}) \lambda_{\textrm{min}}(\bm{C})$ \cite{Wan93}, and, since $\lambda(\bm{P}) \leq 1$, it holds that $\lambda_{\textrm{min}}(\bm{P}) \lambda_{\textrm{min}}(\bm{C}) \leq \lambda_{\textrm{min}}(\bm{C})$. Thus, $\Delta \leq \lambda_{\textrm{min}}(\bm{P}) \lambda_{\textrm{min}}(\bm{C})$ with $\lambda_{\textrm{min}}(\bm{P}) = \nu$ satisfies both conditions. 

\bibliographystyle{IEEEtran}
\bibliography{IEEEabrv,References}

\end{document}